\documentclass[12pt]{article}

\usepackage{amsmath,amssymb,epsfig,subfigure,color,soul,fleqn,rotating}
\usepackage{ulem}
\usepackage{cite}
\usepackage[colorlinks,citecolor=blue,urlcolor=blue,linkcolor=blue]{hyperref}

\newcommand{\be}{\begin{eqnarray}}
\newcommand{\ee}{\end{eqnarray}}
\newcommand{\nee}{\nonumber\end{eqnarray}}

\newcommand{\drbar}{{\overline{\rm DR}}}

\newcommand{\mnt}[1] {m_{\ti \x^0_{#1}}}

\newcommand{\msg}    {m_{\ti g}}
\newcommand{\msu}[1] {m_{\ti u_{#1}}}

\def\gev             {{\rm GeV}}

\def\be            {\begin{equation}}
\def\ee            {\end{equation}}
\def\bea            {\begin{eqnarray}}
\def\eea            {\end{eqnarray}}

\def\a              {\alpha}
\def\b               {\beta}
\def\d               {\delta}

\def\x               {\chi}
\def\ti              {\tilde}

\def\st              {\ti t}

\def\sg              {\ti g}

\def\su                {\ti{u}}
\def\sc                {\ti{c}}
\def\st                {\ti{t}}
\def\sto                  {\ti{t}}
\def \sca                 {\ti{c}}

\def\dll            {\d^{LL}_{23}}
\def\durr            {\d^{uRR}_{23}}
\def\durl            {\d^{uRL}_{23}}
\def\dulr            {\d^{uLR}_{23}}

\newcommand{\AddrVienna}{
\it Universit\"at Wien, Fakult\"at f\"ur Physik,
A-1090 Vienna, Austria \\}

\newcommand{\AddrGAKUGEI}{%
 \it Department of Physics, Tokyo Gakugei University, Koganei,
Tokyo 184-8501, Japan\\}

\newcommand{\AddrRIKEN}{%
 \it RIKEN Nishina Center for Accelerator-Based Science, 
 Wako, Saitama 351-0198, Japan\\}

\newcommand{\AddrHEPHY}{%
 \it Institut f\"ur Hochenergiephysik der \"Osterreichischen Akademie
der Wissenschaften, A-1050 Vienna, Austria\\}

\title{Higgs boson decay to charm pair at full one-loop level in the MSSM with flavour violation}

\author{H. Eberl${}^{1\footnote{Talk presented at the International Workshop on Future Linear Colliders (LCWS14), Belgrade, \mbox{Serbia, 6-10 October 2014.}}}$, A. Bartl${}^{2}$, E. Ginina${}^{1}$, K.~Hidaka${}^{3,4}$, W.~Majerotto${}^1$}
\date{
\small $^1$ \AddrHEPHY 
          $^2$ \AddrVienna        
         $^3$ \AddrGAKUGEI 
         $^4$ \AddrRIKEN}

\definecolor{darkgreen}{rgb}{0,.5,0}

\begin{document}

\begin{flushright}
HEPHY-PUB 947/14\\
RIKEN-MP-102\\
UWThPh-2014-35
\end{flushright}
\begingroup
\let\newpage\relax
\maketitle
\endgroup

\maketitle
\thispagestyle{empty}

\begin{abstract}
We study the decay of the lightest neutral Higgs boson to a charm quark pair at full one-loop level 
in the MSSM with non-minimal quark flavour violation (QFV). In the numerical analysis we consider mixing between the second and the third squark generation and all relevant constraints from B meson data are taken into account. It is shown that the full one-loop corrected decay width can be quite sensitive to the MSSM QFV parameters due to large $\tilde c - \tilde t$ mixing and large trilinear couplings. After summarising the theoretical and experimental errors,
we conclude that an observation of these SUSY QFV effects is possible at the ILC.

%
%

%
\end{abstract}

\clearpage

\section{Introduction}

After having found the Higgs boson at the LHC it is a logical next step to measure its properties. With the 
next run of LHC starting in 2015 at higher energy ($\sqrt{s}=14$ TeV) and higher luminosity, more precise data on 
Higgs boson observables will be available. Even more precise data can be expected at a future $e^+ e^-$ linear collider (ILC).
The discovered Higgs boson could be the lightest neutral Higgs boson $h^0$ of the 
Minimal Supersymmetric Standard Model (MSSM)~\cite{LHCcrosssecs, Djouadi:2005gi}. 
Quark flavour violation (QFV) in the squark sector may change the decay widths of $h^0$ at one-loop level
which leads to a deviation of the SM prediction, especially because of the mixing
between the second and the third squark generations 
($\tilde{c}_{L,R}-\st_{L,R}$ mixing). In this conference contribution we will focus on the decay $h^0 \to c \bar c$. Not all details
will be shown in this work, but they can be found in~\cite{Bartl:2014bka}. 

\section{Squark generation mixing in the MSSM}

The mass matrix of the up squarks can be written in the super-CKM basis of 
$\su_{0 \alpha} =
(\su_{{\rm L}}, \sc_{{\rm L}}, \st_{{\rm L}}, \su_{{\rm R}}, \sc_{{\rm R}}, \st_{{\rm R}})$
in its most general $3 \times 3$~block form as the hermitian matrix
\begin{equation}
    {\cal M}^2_{\tilde{u}} = \left( \begin{array}{cc}
        {\cal M}^2_{\tilde{u},LL} & {\cal M}^2_{\tilde{u},LR} \\[2mm]
        {\cal M}^2_{\tilde{u},RL} & {\cal M}^2_{\tilde{u},RR} \end{array} \right), \quad
 {\rm with} \quad
 \begin{array}{l}
    {\cal M}^2_{\tilde{u},LL} = V_{\rm CKM} M_Q^2 V_{\rm CKM}^{\dag} + D_{\tilde{u},LL}{\bf 1} + \hat{m}^2_u\, ,\\[2mm]
    {\cal M}^2_{\tilde{u},RR} = M_U^2 + D_{\tilde{u},RR}{\bf 1} + \hat{m}^2_u\, , \\[2mm]
    {\cal M}^2_{\tilde{u},RL} = {\cal M}^{2\dag}_{\tilde{u},LR} =
      \frac{v_2}{\sqrt{2}} T_U - \mu^* \hat{m}_u\cot\beta\, ,
 \end{array}             
 \label{EqMassMatrix}
\end{equation}
with the soft SUSY breaking $3 \times 3$ mass matrices $M_Q^2, M_U^2$, and $T_U$, 
$\hat{m}_{u}$ denotes the diagonal up-type mass matrix, $D_{\tilde{u},LL}$ and $D_{\tilde{u},RR}$ are D-term
contributions and $V_{\rm CKM}$ is the CKM matrix,
$\mu$ the higgsino mass parameter, and $\tan\beta$ the ratio of the vacuum expectation values of the neutral Higgs fields $v_2/v_1$, with $v_{1,2}=\sqrt{2} \left\langle H^0_{1,2} \right\rangle$.\\
After diagonalization with the rotation matrix $U^{\tilde{u}}_{6 \times 6}$, the mass eigenstates are obtained:
$ \tilde{u}_i=U^{\tilde{u}}_{i\alpha} \tilde{u}_{0 \alpha}, ~ $where$ ~U^{\tilde{u}} {\cal M}_{\tilde u}^2U^{\tilde{u}\dagger} = {\rm diag} (m^2_{\tilde{u}_1},...,m^2_{\tilde{u}_6})$, $ ~ m_{\tilde{u} _i} < m_{\tilde{u}_j} ~ {\rm for} ~ i<j $. 
The formulas for the sdown system are analogous~\cite{Bartl:2014bka}.\\

In order to avoid the strong experimental constraints from Kaon physics, in our numerical 
discussion we only allow mixing between the second and third squark generation.  This is encoded in the dimensionless
parameters
\begin{eqnarray}
\delta^{LL}_{23} & \equiv & M^2_{Q 23} / \sqrt{M^2_{Q 22} M^2_{Q 33}}\quad \sim \sc_L - \st_L\, {\rm mixing}\, ,
\label{eq:InsLL}\\[1mm]
\delta^{uRR}_{23} &\equiv& M^2_{U 23} / \sqrt{M^2_{U 22} M^2_{U 33}}\quad \sim \sc_R - \st_R\, {\rm mixing}\, ,
\label{eq:InsRR}\\[1mm]
\delta^{uRL}_{23} &\equiv& (v_2/\sqrt{2} ) T_{U 23}/ \sqrt{M^2_{U 22} M^2_{Q 33}}\quad \sim \sc_R - \st_L\, {\rm mixing}\, ,
\label{eq:InsRL}\\[1mm]
\delta^{uLR}_{23} &\equiv& (v_2/\sqrt{2} ) T_{U 32}/ \sqrt{M^2_{U 33} M^2_{Q 22}}\quad \sim \sc_L - \st_R\, {\rm mixing}\, .
\label{eq:InsLR}
\end{eqnarray}

\section{Constraints on the MSSM parameters}
\label{sec:constr}

In the numerical analysis theoretical constraints from vacuum stability on the $T_U$ matrix are considered.
On the experimental side the SUSY mass limits from direct collider searches and the measured Higgs mass
including the theoretical uncertainty from SUSY, $m_{h^0} = (125.15 \pm 2.48)$~GeV,
are taken into account. Furthermore, the constraints from the electroweak precision observable $\rho$  
and relevant low-energy measurements are checked, using
\begin{eqnarray}
  \Delta \rho~({\rm SUSY}) & < &  0.0012\, , \nonumber \\
   {\rm B}(b \to s \gamma) &=& (3.4 \pm 0.61) \times 10^{-4}\, , \nonumber \\
   \Delta M_{B_s} &=& (17.77 \pm 3.3)~{\rm ps}^{-1}\, , \nonumber \\
  {\rm B} (b\to s~\mu^+\mu^-) &=& (1.60 \pm 0.91) \times 10^{-6}\,, \nonumber \\ 
  {\rm B} (B_s\to \mu^+\mu^-) &=& (2.9 \pm 1.44) \times 10^{-9}\, .\nonumber 
\end{eqnarray}

\section{{\boldmath $h^0 \to c \bar c$} @ full one-loop level}

We study the decay of the lightest neutral Higgs boson, $h^0 \to c \bar c$,
at full one-loop level in the MSSM with non-minimal flavour violation and real input parameters.
The renormalisation is done in the $\overline{\rm DR}$ renormalisation scheme. 
The partial decay width, including one-loop contributions can be written as
\begin{equation}
\Gamma(h^0 \to c \bar{c})=\Gamma^{\rm tree}(h^0 \to c \bar{c})+ \sum_{x=g, \ti{g}, {\rm EW}} \d \Gamma^x \, ,
\label{decaywidth}
\end{equation}
where the tree-level decay width is
\begin{equation}
\Gamma^{\rm tree}(h^0 \to c \bar{c})=\frac{3}{8 \pi} m_{h^0} (s_1^c)^2 \beta^{3}\,, 
\label{decaywidttree}
\end{equation}
with $\beta = \sqrt{ 1- \frac{4 m_c^2}{m_{h^0}^2}}$, $m_{h^0}$ is taken on-shell and the tree-level coupling $s_1^c$  is the $\overline{\rm DR}$ running 
one given at the scale $Q = m_{h^0}$. It reads
\begin{equation}
 s_1^c=-g \frac{m_c}{2 m_W} \frac{\cos{\a}}{\sin{\b}} =-\frac{h_c}{\sqrt{2}} \cos{\a}\,,
\label{treecoup}
\end{equation}
where $\alpha$ is the mixing angle of $h^0$ and $H^0$. The renormalised one-loop contributions are
\be
\d \Gamma^{\ti{g}} = \frac{3}{4 \pi} m_{h^0} s_1^c {\rm Re}(\d S_1^{c, \ti{g}}) \beta^{3}\,,
\label{1loopcontr}
\ee
and
\be
\d \Gamma^{g/\rm EW} = \frac{3}{4 \pi} m_{h^0} s_1^c {\rm Re}(\d S_1^{c, g/\rm EW}) \beta^{3}+ \Gamma^{\rm hard}(h^0 \to c \bar{c} g/\gamma)\,.
\label{1loopgEWcontr}
\ee
The renormalised UV finite one-loop amplitude of the process is a sum of all vertex diagrams, the amplitudes arising from the wave-function renormalisation constants and the amplitudes arising from the coupling counter terms. The full one-loop calculation with virtual contribution is done with our own derived
counter terms with the help of FeynArts~\cite{Hahn:2000kx} and FormCalc~\cite{Hahn:1998yk}. \\

For having also an IR convergent result hard photon/gluon radiation is
taken into account.

\subsection{One-loop gluon contribution} 

Summing up all contributions with a virtual gluon in the loop we get\\
$\d \Gamma^{g}=\frac{3}{4 \pi} m_{h^0} s_1^c ~{\rm Re}(\d S_1^{c, g})\b^3$
and the width with the hard gluon radiation can be written as
\be
\Gamma^{\rm hard}(h^0 \to c \bar{c} g)=\frac{3}{8 \pi} m_{h^0} (s_1^c)^2 \b^3  \frac{4}{3}\frac{ \a_s}{\pi} \Delta^{{\rm H}, {\rm hard}} (\b)\;.
\label{delhard}
\ee
Adding both we can write the result as
\be
\Gamma^{g}(h^0 \to c \bar{c})=\Gamma^{\rm tree}(m_c\vert_{\rm OS})\left( 1+\frac{4}{3}\frac{\a_s}{\pi} \Delta^{\rm H} (\b) \right)\,,
\label{resg1}
\ee
where $m_c \vert_{\rm OS}$ denotes the on-shell (OS) charm quark mass and $\Delta^{\rm H} (\b)$ is the result of~\cite{Braaten&Leveille}.
In the following we assume $m_c \ll m_{h^0}$ ($\b \to 1$)
which is a very good approximation. We then have
\be
\Delta^{\rm H} = -3 \ln \frac{m_{h^0}}{m_c\vert_{\rm OS}} +\frac{9}{4}
\ee
The large logarithm $\ln \frac{m_{h^0}}{m_c\vert_{\rm OS}}$ can be absorbed by redefining the charm quark mass~\cite{Eberl:1999he}.
Furthermore we have included in addition the gluonic $\a_s^2$ contributions, taken from~\cite{Spira:1997dg}.
We get 
\be
\Gamma^{g, {\rm impr}}(h^0 \to c \bar{c}) =\Gamma^{\rm tree}(m_c|_{\rm SM})\left( 1+\frac{17}{3}\frac{\a_s}{\pi} + \a_s^2~{\rm contrib.}\right)\
\ee
with $m_c|_{\rm SM}$ the SM $\overline{\rm MS}$ running charm quark mass at the scale $m_{h^0}$.

\subsection{One-loop gluino contribution}

The one-loop gluino contribution to  $\Gamma(h^0 \to c \bar{c})$, renormalised in the $\overline{\rm DR}$-scheme reads
\be
\d \Gamma^{\sg}=\frac{3}{4 \pi}m_{h^0}~ s_1^c ~{\rm Re}(\d S_1^{c, \sg})\b^3\,.
\label{decaywidthgluino}
\ee 
Using the abbreviations
$\a_{ij} = U^{\su *}_{i2}U^{\su}_{j2}+U^{\su *}_{i5}U^{\su}_{j5}$  and $\b_{ij} = U^{\su *}_{i2}U^{\su}_{j5}+U^{\su *}_{i5}U^{\su}_{j2}$,
and $m_c, m_{h^0}$ are much smaller than the gluino and squark masses, we get
\be
\d S_1^{c, \sg} = \frac{\a_s}{3 \pi} \sum_{i,j=1}^{6}\bigg\{ \msg \b_{ij} \left( G_{ij1}^{\su} C_0^{ij}  + 4 s_1^c \d_{ij}  \dot{B}_0^{i}\right)
+ s_1^c  \d_{ij} \left( \a_{ii} B_1^{i} + \Delta\right) \bigg\}\, ,
\label{gluinocontr}
\ee
with $\Delta$ the UV divergence, and 
$B_{k}^i=B_{k}(0,\msg^2, \msu i^2)$, $\dot{B}_{0}^i=\frac{\partial B_{0}(p^2, \msg^2, \msu i^2)}{\partial p^2}\big\vert_{p^2=0}$, and\\
$C_{0}^{ij}=C_{k}(0, 0, 0, \msg^2, \msu i^2,\msu j^2)$ are Passarino-Veltman loop integrals.
$G_{ij1}^{\su}$ denotes the $h^0 \su_i^* \su_j$ coupling.\\
At first sight it seems that the gluino contribution does not decouple for $\msg \to \infty$ because $B_1$ grows with $\ln \frac{\msg^2}{m_{h^0}^2}$. 
However, the tree-level coupling $s_1^c$ (eq.~(\ref{treecoup})) contains a factor $m_c$, 
$m_c (m_{h^0}) \vert_{\drbar} = m_c (m_c) \vert_{\overline{\rm MS}} +\d m_c^{\tilde{g}}+\ldots$. 
Thus the sum $\Gamma^{\rm tree}+ \d \Gamma^{\tilde{g}}$ is indeed decoupling for $\msg \to \infty$. Analogously, this also holds for the chargino and neutralino contributions.

\section{Numerical results}
\label{sec:num}

In order to demonstrate clearly the effect of QFV in the MSSM, we have explicitly chosen a reference scenario with a rather strong $\ti{c}-\ti{t}$ mixing. The MSSM parameters at $Q = 125.5~\gev \simeq m_{h^0}$ are given in Table~\ref{basicparam}. 
This scenario satisfies all present experimental and theoretical constraints 
discussed in Section~\ref{sec:constr}. For calculating the masses and the mixing, as well as the low-energy observables, especially 
those in the B meson sector, we use the public code 
SPheno v3.3.3~\cite{SPheno1, SPheno2}. 
%
\begin{table}[h!]
\caption{Reference QFV scenario: the basic MSSM $\overline{DR}$ parameters at $Q = 125.5~{\rm GeV} \simeq m_{h^0}$,
except for $m_{A^0}$ which is the pole mass, 
with $T_{U33} = - 2050$~GeV. All other squark parameters not shown here are zero. }
\begin{center}
\begin{tabular}{|c|c|c|c|c|c|}
  \hline
 $M_1$ & $M_2$ & $M_3$ & $\mu$ & $\tan \beta$ & $m_{A^0}$\\
 \hline \hline
 250~\gev  &  500~\gev &  1500~\gev& 2000~\gev & 20 &  1500~\gev \\
  \hline
\end{tabular}
\vskip0.4cm
\begin{tabular}{|c|c|c|c|}
  \hline
   & $\alpha = 1$ & $\alpha= 2$ & $\alpha = 3$ \\
  \hline \hline
   $M_{Q \alpha \alpha}^2$ & $(2400)^2~\gev^2$ &  $(2360)^2~\gev^2$  & $(1850)^2~\gev^2$ \\
   \hline
   $M_{U \alpha \alpha}^2$ & $(2380)^2~\gev^2$ & $(1050)^2~\gev^2$ & $(950)^2~\gev^2$ \\
   \hline
   $M_{D \alpha \alpha}^2$ & $(2380)^2~\gev^2$ & $(2340)^2~\gev^2$ &  $(2300)^2~\gev^2$  \\
   \hline
\end{tabular}
\vskip0.4cm
\begin{tabular}{|c|c|c|c|}
  \hline
   $\delta^{LL}_{23}$ & $\delta^{uRR}_{23}$  &  $\delta^{uRL}_{23}$ & $\delta^{uLR}_{23}$\\
  \hline \hline
   0.05 & 0.2 &  0.03   &  0.06  \\
    \hline
\end{tabular}
\end{center}
\label{basicparam}
\end{table}

\noindent
We get for this scenario  $\mnt{1} = 260$~GeV, $m_{h^0} = 126.1$~GeV, $\msg = 1473$~GeV,
$\msu{1} = 756$~GeV, and $\msu{2} = 965$~GeV. 

\clearpage
\noindent
The squared coefficients of the flavour decomposition of the two 
lighter squarks $\su_1$ and $\su_2$ are
\begin{center}
\begin{tabular}{|c|c|c|c|c|c|c|c|}
  \hline
  & $\su_L$ & $\sca_L$ & $\sto_L$ & $\su_R$ & $\sca_R$ & $\sto_R$ \\
  \hline \hline
 $\su_1$  & $0$ & $0.0004$ & $0.012$ & $0$ & $0.519$ & $0.468$ \\
  \hline 
  $\su_2$  & $0$ & $0.0004$ & $0.009$ & $0$ & $0.480$ & $0.509$ \\
  \hline
\end{tabular}
\end{center}

\begin{figure*}[h!]
\centering
\subfigure[]{
   { \mbox{\hspace*{-1cm} \resizebox{8.cm}{!}{\includegraphics{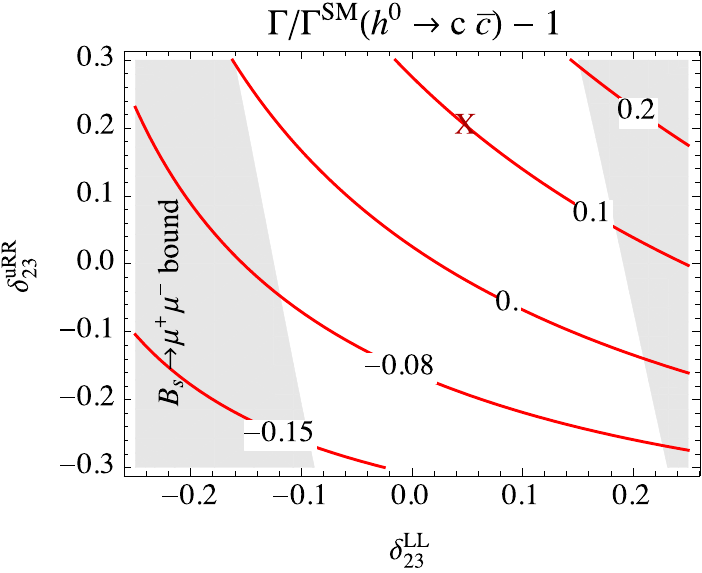}} \hspace*{-0.8cm}}}
   \label{fig1a}}
 \subfigure[]{
   { \mbox{\hspace*{+0.cm} \resizebox{8.cm}{!}{\includegraphics{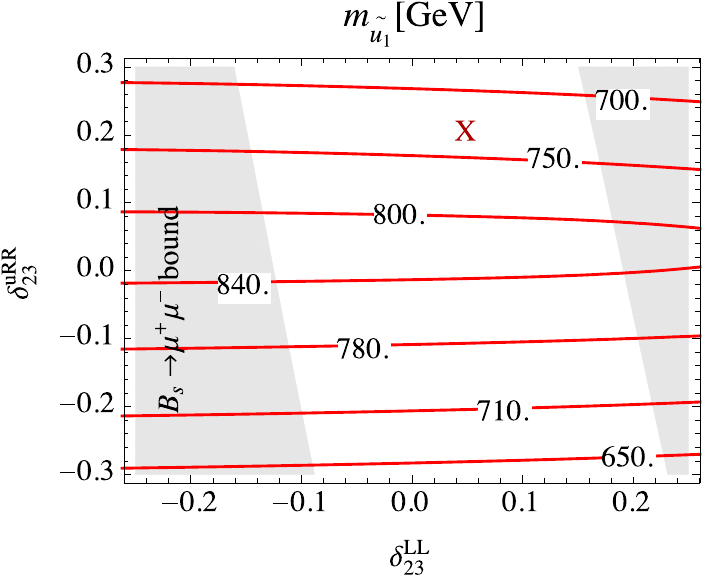}} \hspace*{-1cm}}}
  \label{fig1b}}\\
\caption{Dependence on the QFV parameters $\dll$ and $\durr$ of (a) $\Gamma(h^0 \to c \bar{c})$/$\Gamma^{\rm SM}(h^0 \to c \bar{c})$ 
and (b) the mass of the lightest  squark $\su_1$ in GeV. The gray region is excluded by the constraint from the B($B_s \to \mu^+ \mu^-$) data.
}
\label{fig1}
\end{figure*}

In Fig.~\ref{fig1a}  we show the deviation of the $\Gamma(h^0 \to c \bar{c})$ 
from the SM width $\Gamma^{\rm SM}(h^0 \to c \bar{c})= 0.118$ MeV~\cite{pdg2014}. 
This deviation varies between -15\% and 20\%. It is interesting to mention that we
obtain $\Gamma^{\rm QFC}(h^0 \to c \bar{c})= 0.116$ MeV for the full one-loop width 
in the QFC MSSM case for our reference scenario corresponding to Table~\ref{basicparam}. 
This means that the QFC supersymmetric contributions  change the width 
$\Gamma(h^0 \to c \bar{c})$ by only \mbox{$\sim $ -1.5\%} compared to  the SM value.
Comparing our non-improved QFC one-loop result with 
FeynHiggs-2.10.2~\cite{Heinemeyer:1998yj} we have a difference less than 1\%. 
The mass of the lightest squark $\su_1$ can vary in the region $-0.3<\durr<0.3$  between 650 GeV and 850 GeV, as seen in Fig.~\ref{fig1b}.
Note, that in principle a larger region of $\durr$ is possible because there is no experimental restriction applicable up to now 
but the trivial bound $\msu{1} > \mnt{1}$.

\begin{figure*}[h!]
\centering
\subfigure[]{
   { \mbox{\hspace*{-1cm} \resizebox{8.cm}{!}{\includegraphics{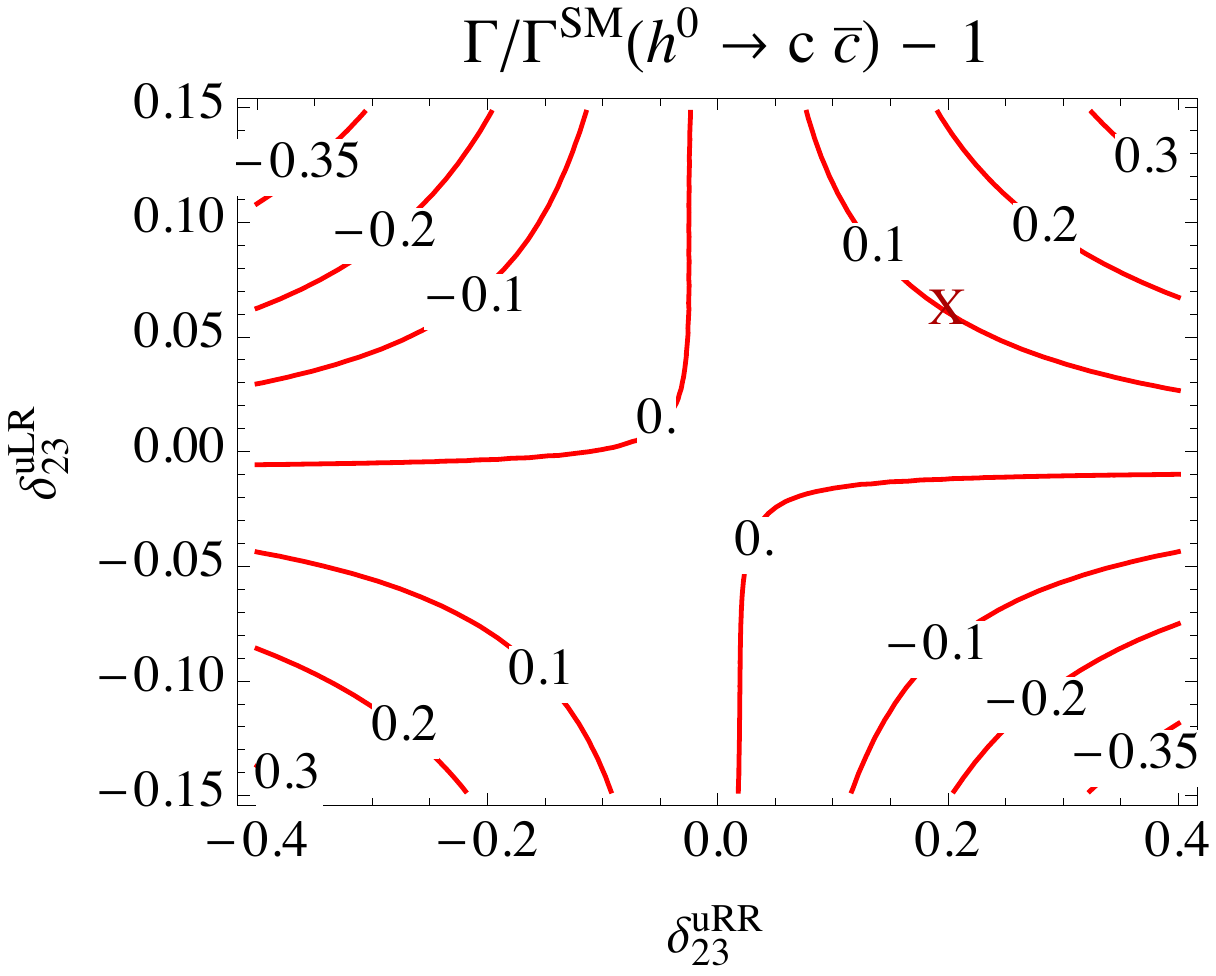}} \hspace*{-0.8cm}}}
   \label{fig2a}}
 \subfigure[]{
   { \mbox{\hspace*{+0.cm} \resizebox{8.cm}{!}{\includegraphics{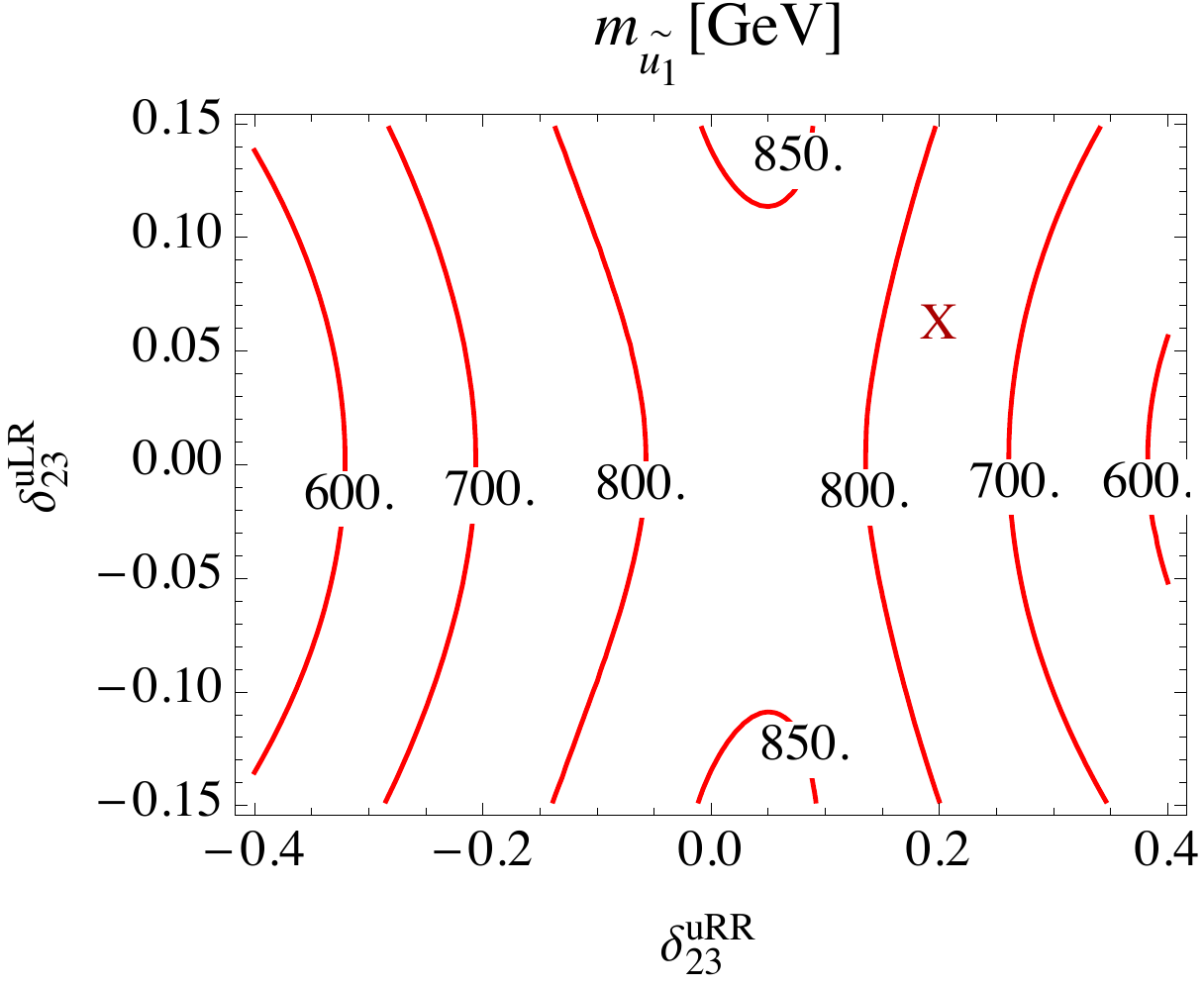}} \hspace*{-1cm}}}
  \label{fig2b}}\\
\caption{Dependence on the QFV parameters $\durr$ and $\dulr$ of (a) $\Gamma(h^0 \to c \bar{c})$/$\Gamma^{\rm SM}(h^0 \to c \bar{c})$ 
and (b) the mass of the lightest squark $\su_1$ in GeV.} 
\label{fig2}
\end{figure*}

In Fig.~\ref{fig2} we show the corresponding plots for the dependences on the QFV parameters $\durr$ and $\dulr$. As seen in Fig.~\ref{fig2a},  
the deviation from the SM value $\Gamma^{\rm SM}(h^0 \to c \bar{c})$ is between -35\% and 30\%. The mass of $\su_1$ varies between 600 GeV and 850 GeV, as seen in Fig.~\ref{fig2b}. Fig.~\ref{fig2a} shows a symmetry around the origin. In eq.~(\ref{gluinocontr}) the term with
$\msg \b_{ij} = \msg (U^{\su *}_{i2}U^{\su}_{j5}+U^{\su *}_{i5}U^{\su}_{j2})$ which has terms $\propto \durr \dulr$ and
$\propto  \dll \durl$ 
can become numerically large. We checked that the contour plot on $\dll$ and $\durl$ indeed shows the same symmetry, but the
phenomenologically allowed window is much smaller there.
 
The strong dependences of the width $\Gamma(h^0 \to c \bar{c})$ on the QFV parameters 
shown in this section can be explained as follows.
First of all, the scenario chosen is characterised by large QFV parameters, which in our case are the large $\ti{c}_{L,R}-\ti{t}_{L,R}$ 
mixing parameters $\dll, \durr, \durl , \dulr$, 
and particularly large QFV trilinear couplings $T_{U23}, T_{U32}$ (Note that  $\durl \sim T_{U23}$ and $\dulr \sim T_{U32}$).    
In such a scenario, the lightest up-type squarks $\su_{1,2}$ are strong 
    admixtures of $\ti{c}_{L,R}-\ti{t}_{L,R}$, and, hence, the couplings 
    $\su_{1,2} \su_{1,2}^* h^0(\sim {\rm Re}(H_2^0))$ in the vertex graph with a gluino are strongly 
    enhanced.
In addition, large $\ti{t}_{L}-\ti{t}_{R}$ mixing due to the large QFC trilinear couplings $T_{U33}$ occurs. 

\section{Theoretical error estimation}
There are two uncertainties in the theoretical prediction of the width. The first is the scale uncertainty which gives an estimate on the size of
higher loop contributions. At our reference point we get $\sim 0.5\%$ for that. 
The other one is the so called parametric uncertainty due to the errors of the SM input
parameters, for our studied process only $m_c(m_c)\vert_{\overline{\rm MS}} = 1.275$~GeV with $\d m_c/m_c = 2\%$ and 
$\alpha_s(m_Z) \vert_{\overline{\rm MS}} = 0.1185$ with $\d \a_s/\a_s = 0.5\%$ are relevant. At our reference point we get $5.2\% \oplus 2\%$.
For the total error in the width at our reference point we estimate
\bea
\sqrt{5.2\%^2+2\%^2}+0.5\%&\approx& 6.1\%\,,
\label{ouruncert}
\eea

As seen in Section~\ref{sec:num}, the deviation $\Gamma(h^0 \to c \bar{c})/\Gamma^{\rm SM}(h^0 \to c \bar{c})$ can be as large as $\sim \pm35$\%. 
Such a large deviation can be observed at ILC (500 GeV) with 1600 (500) ${\rm fb}^{-1}$, 
where the expected experimental error in the width is 3\% (5.6\%)~\cite{Asner:2013psa, Tian@EPS-HEP2013} .
A measurement of $\Gamma(h^0 \to c \bar{c})$ at LHC (even with the high luminosity upgrade) is demanding due to uncertainties in the charm-tagging.

\section{Conclusion}

We have calculated the width $\Gamma (h^0 \to c \bar{c})$  at full one-loop level within the MSSM with quark flavour violation. In particular, we have studied $\ti{c}_{R,L}-\ti{t}_{R,L}$ mixing, taking into account the experimental constraints from B-physics, $m_{h^0}$ and SUSY particle searches. The width $\Gamma (h \to c \bar{c})$ turns out to be very sensitive to $\ti{c}_{R,L}-\ti{t}_{R,L}$ mixing.

In our calculation we have used the $\drbar$ renormalisation scheme. In particular, we have derived the explicit formula for the dominant gluino loop contribution. We also have performed a detailed numerical study of the QFV 
parameter dependence of the width. Whereas the width $\Gamma (h^0 \to c \bar{c})$ 
in the QFC MSSM case is only slightly different from its SM value, in the QFV 
case this width can deviate from the SM by up to $\sim \pm 35\%$.

We have estimated the theoretical uncertainties of 
$\Gamma (h^0 \to c \bar{c})$ and have shown that the SUSY QFV contribution 
to this width can be observed at the ILC.

%
\section*{Acknowledgments}

This work is supported by the "Fonds zur F\"orderung der wissenschaftlichen Forschung (FWF)" of Austria, project No. P26338-N27.
K.~H. thanks Prof. Koji Hashimoto for the warm hospitality at RIKEN Nishina Center for Accelerator-Based Science.

\end{document}